\documentclass[aps,prb,reprint,twocolumn,groupedaddress,superscriptaddress,showpacs,floatfix,nobalancelastpage]{revtex4-1}
\usepackage{graphicx}

\def \musr {$\mu^+$SR}
\def \fmu {F--$\mu$}
\def \fmuf {F--$\mu$--F}

\begin{document}

\title{Quantum states of muons in fluorides}

\author{J.~S.~M\"{o}ller}
\email{johannes.moeller@physics.ox.ac.uk}
\affiliation{Department of Physics, Clarendon Laboratory, Oxford University, Parks Road, Oxford, OX1 3PU, UK}

\author{D.~Ceresoli} 
\affiliation{Istituto di Scienze e Tecnologie Molecolari CNR, via Golgi 19, 20133 Milano, Italy}

\author{T.~Lancaster}
\affiliation{Centre for Materials Physics, Durham University, South Road, Durham, DH1 3LE, UK}

\author{N.~Marzari}
\affiliation{Theory and Simulation of Materials (THEOS), \'{E}cole Polytechnique F\'ed\'erale de Lausanne, 1015 Lausanne, Switzerland}

\author{S.~J.~Blundell}
\affiliation{Department of Physics, Clarendon Laboratory, Oxford University, Parks Road, Oxford, OX1 3PU, UK}

\date{March 18, 2013}

\begin{abstract}
Muon-spin relaxation (\musr) is a sensitive probe of magnetism, but its utility can be severely limited by the lack of knowledge of the muon implantation site and the extent to which the muon perturbs its host. We demonstrate systematically that these problems can be addressed accurately using electronic-structure calculations. We show that diamagnetic muons introduce significant short-ranged distortions in ionic insulators that would lead to systematic errors on magnetic moments determined by \musr, and quantify these. The \fmuf\ complex formed by muons in many fluorides can be understood as an exotic molecule-in-a-crystal defect with a zero-point energy larger than that of any naturally-occurring triatomic molecule.
\end{abstract}


\maketitle

Muon-spin relaxation (\musr) involves implanting spin-polarized positive muons in a sample in order to probe the local magnetic structure.\cite{Blundell99} \musr\ is an extremely sensitive probe of magnetism\cite{Pratt2011Nature} but has two significant limitations. The first concerns the lack of knowledge of the site of implantation of the muon, which hinders the measurement of magnetic moments using \musr. Second, the unknown extent of the perturbation due to the muon of the local crystal and electronic structure of the host has been the cause for increased concern since \musr\ is frequently employed in the study of systems that lie on the verge of ordering\cite{Kojima1995PRL,Pratt2011Nature,Lancaster2006PRB,Blundell1997JPhysCM} or where doping is a critical parameter.\cite{Amit2010PRB,Luke1990PRB,Jack2012PRB} Previous first-principles studies have focussed on the paramagnetic states formed by muons and protons in semiconductors.~\cite{VdWZnO,Porter1999,Luchsinger1997} Diamagnetic muon states (where the contact hyperfine coupling is negligible) have received considerably less attention, in spite of their greater utility in the study of magnetic materials. Here we present a detailed microscopic study based on density-functional theory~(DFT) of the dia- and paramagnetic muon states in a series of fluorides, where detailed information about the geometry of the diamagnetic muon site is experimentally accessible, enabling an accurate comparison with first-principles predictions. 
 
In host compounds containing fluorine, diamagnetic muons can couple strongly to the fluoride ions often forming linear \fmuf\ complexes,~\cite{brewer86PRB} although bent \fmuf\ and \fmu\ geometries have been shown to exist as well.~\cite{Lancaster2007PRL} The magnetic dipolar coupling between muon and fluorine nuclear spin gives rise to a signal that is sensitive to the geometry of the muon-fluorine state, allowing an accurate experimental determination of the muon's local site geometry.~\cite{brewer86PRB,Lancaster2007PRL} In the series of non-magnetic ionic insulators LiF and NaF (rock-salt structure, $a=4.03$~\AA\ and 4.78~\AA), CaF$_2$ and BaF$_2$ (fluorite structure, $a=5.46$~\AA\ and 6.20~\AA), and for the antiferromagnetic (AFM) insulator CoF$_2$ (rutile-type structure, $a=4.70$~\AA\ and $c=3.18$~\AA), we demonstrate the high accuracy with which quantitative information about the muon both in the dia- and the paramagnetic state can be obtained with DFT. We show that diamagnetic muons cause significant short-ranged perturbations of the host, not limited to the fluorides bound in the \fmuf\ state (indeed the distortion of the neighbouring cations can exceed that of the fluorides). This introduces systematic errors on magnetic moments in ionic insulators determined by \musr, which we quantify. We study the quantum behavior of the muon and the heavier proton in both charge states. 

The {\it ab initio} calculations were performed with the {\sc Quantum ESPRESSO} package.~\cite{qespresso} Unless indicated otherwise, calculations were performed in a supercell containing $2\times2\times2$ conventional unit cells. The charge state of the muon was determined by the charge of the supercell (+1 for diamagnetic and neutral for paramagnetic states). A muon was placed in several randomly chosen low-symmetry sites and all ions were allowed to relax until the forces on all ions and the energy change had fallen below a convergence threshold.~\cite{supplemental}

\begin{figure}[htbp]
\includegraphics[width=0.4\textwidth]{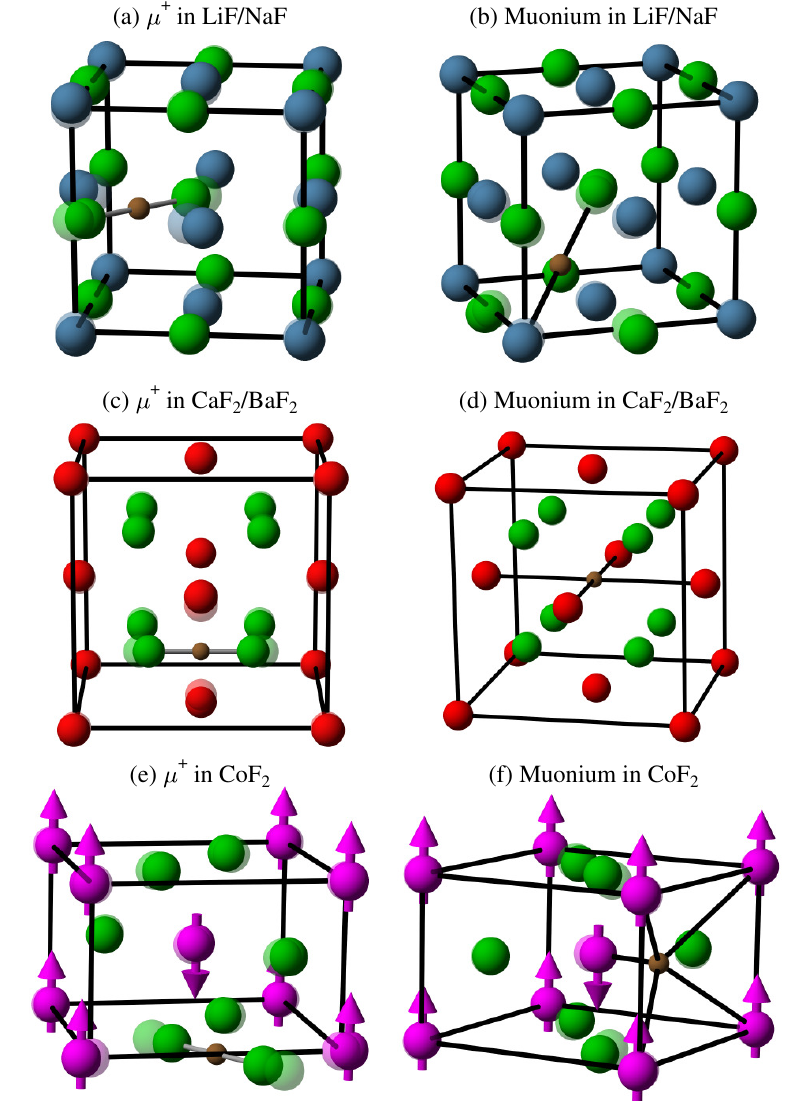}
\caption{\label{fig:structures}(color online). Calculated equilibrium geometries of dia- and paramagnetic muon states in LiF/NaF (Li/Na blue, F green), CaF$_2$/BaF$_2$ (Ca/Ba red), and CoF$_2$ (Co magenta). Translucent spheres represent the equilibrium ionic positions before the muon (brown) is introduced into the crystal. Black lines are a guide to the eye. The $c$~axis is vertical.}
\end{figure}

We first discuss the equilibrium geometries of the muon states obtained from our calculations, shown in Fig.~\ref{fig:structures}. Our results predict the formation of diamagnetic linear \fmuf\ states in all of the compounds of this series. This is in agreement with previous experimental data for non-magnetic LiF, NaF, CaF$_2$, and BaF$_2$.\cite{brewer86PRB} The calculated and experimentally measured bond lengths are tabulated in table~\ref{tab:fmuf}. Our calculated bond lengths are in all cases within 3\% of the experimental values demonstrating the high level of accuracy of these results. We have found no evidence for any other stable diamagnetic states in this series. For CoF$_2$ a detailed experimental study~\cite{Renzi1984Location} has determined the muon site to be the octahedral $\frac{1}{2}00$ site, in agreement with our calculations, although no experimental reports of an \fmuf\ state exist. In that experiment,\cite{Renzi1984Location,Renzi1984Magnetization} the paramagnetic region, where a potential \fmuf\ signal would be present, was studied only under an applied magnetic field, impeding the observation of a potential \fmuf\ signal. Following our first-principles results, we have therefore searched for experimental signatures of an \fmuf\ state in CoF$_2$ in zero applied field. We have found unambiguous experimental evidence\cite{supplemental} for a symmetric, linear \fmuf\ state with a fluoride-fluoride separation of 2.43(2)~\AA, in good agreement with our calculated value of 2.36~\AA. These results demonstrate that DFT is a powerful tool for determining diamagnetic muon sites. 

\begin{figure}[htbp]
\includegraphics[width=0.44\textwidth]{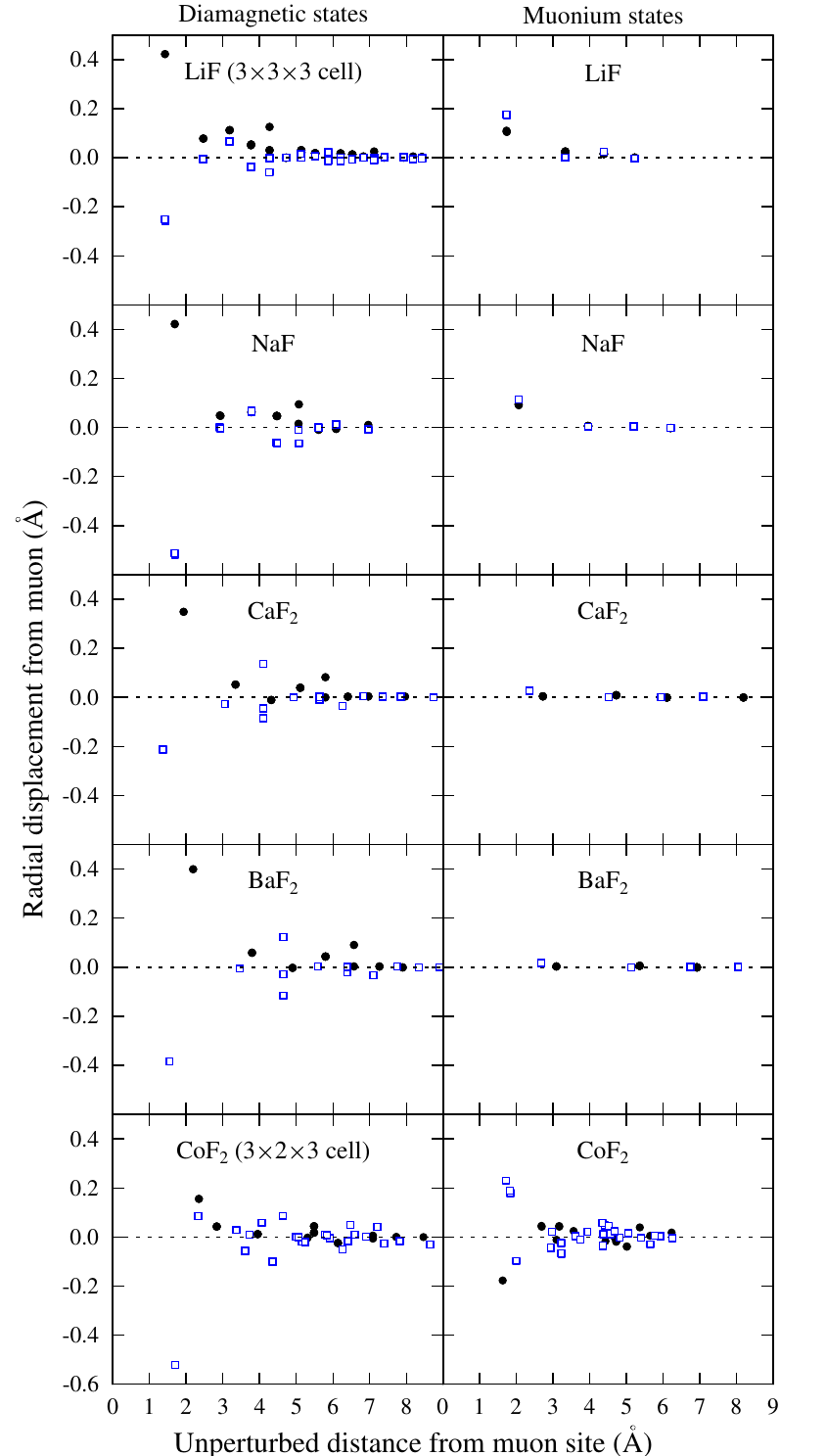}
\caption{\label{fig:distortions}(color online). The radial displacements of the ions as a function of their distance from the muon site in the unperturbed crystal. Cation (black circle), fluoride (blue square). For all compounds the displacements are well converged already on a $2\times2\times2$ supercell but for LiF and CoF$_2$ the \fmuf\ displacements are shown for larger cells.}
\end{figure}

Using a supercell approach, we can also quantify the extent of the perturbation of the implanted muon on its host. The calculated structures (Fig.~\ref{fig:structures}) allow us to study the radial displacements of the ions as a function of their unperturbed distance from the muon site (Fig.~\ref{fig:distortions}). The calculated displacements demonstrate that the muon's perturbation is large but short ranged. While it is known that the perturbation of the fluoride ions must be significant based on the experimentally measured \fmu\ bond lengths of the \fmuf\ states found in many fluorides,\cite{brewer86PRB,Lancaster2007PRL} we can now quantify the perturbation of the cations as well. Since localized magnetic moments would be located on the cation, the cation displacements are particularly pertinent to understanding the effect of the muon's perturbation on experimentally measured \musr\ spectra discussed below. Our results show that in LiF, CaF$_2$, and BaF$_2$ the perturbation of the nearest neighbour (n.n.) cations even exceeds those of the fluoride ions bound in the \fmuf\ state. At short distances the direct Coulomb interaction between the muon and the surrounding ions dominates over the elastic interaction transmitted through the lattice. We therefore expect similar distortions to be present in \emph{any} ionic insulator, regardless of whether it contains fluoride ions or not. Indeed we believe that the formation of the \fmuf\ state somewhat mitigates the n.n.\ cation distortions due to the attraction of negative charge density towards the muon. At short distances all displacements are radially in- or outwards from the muon due to the symmetry of the site. Beyond the n.n.\ shell, elastic interactions cause some non-radial displacement. 

\begin{table}[htbp] %add [H] placement to break table across pages
\begin{ruledtabular}
\begin{tabular}{l | l l l l l l l l} % this means left separator left left left ... 
 & 2r$_{\rm DFT}$ & 2r$_{\rm exp}$ & $\nu_{\rm SS}$ & $\nu_{\rm B}$ & $\nu_{\rm B}$ & $\nu_{\rm AS}$ & ZPE \\ \hline
(FHF)$^{-{\rm a}}$ & 2.36 & 2.28 & 581 & 1289 & 1289 & 1611 & 0.30 \\
(FHF)$^{-{\rm b}}$ & -- & 2.28 & 583 & 1286 & 1286 & 1331$^{\rm c}$ & 0.28 \\
(\fmuf)$^-$ & 2.36 & -- & 581 & 3797 & 3797 & 4748 & 0.80 \\
LiF & 2.34\cite{Bernardini_note} & 2.36(2)\cite{brewer86PRB} & -- & 2825 & 4603 & 4881 & 0.76 \\
NaF & 2.35 & 2.38(1)\cite{brewer86PRB} & -- & 3071 & 4363 & 4813 & 0.76 \\
CaF$_2$ & 2.31 & 2.34(2)\cite{brewer86PRB} & 649 & 2737 & 4481 & 5446 & 0.83 \\
BaF$_2$ & 2.33 & 2.37(2)\cite{brewer86PRB} & 613 & 3033 & 4130 & 4974 & 0.79 \\
CoF$_2$ & 2.36 & 2.43(2) & 585 & 3076 & 3473 & 4570 & 0.73 \\
\end{tabular}
\end{ruledtabular}
\caption{\label{tab:fmuf} Calculated (DFT) and experimental (exp) properties of the diamagnetic \fmuf\ states in solid and vacuum, and of the (FHF)$^-$ molecular ion in vacuum. $r$ (\AA) is the muon-fluoride bond length, $\nu$ is the frequency (cm$^{-1}$) of the symmetric stretch (SS), asymmetric stretch (AS), and bending (B) mode, and ZPE is the zero-point energy (eV). $^{\rm a}$Our calculation. $^{\rm b}$Experimental data (Ref. \onlinecite{Kawaguchi1987}). $^{\rm c}$Ref.~\onlinecite{Hunt1987} reports 1377 cm$^{-1}$.}
\end{table}

We have also investigated the effect that the muon has on the magnetic moments of the surrounding ions. In this series only CoF$_2$ is magnetic. The spin-only moment was estimated from a L\"owdin population analysis to be $2.68\mu_{\rm B}$ per Co ion. This compares with a total moment of 2.60(4)$\mu_{\rm B}$ measured with powder neutron diffraction\cite{Jauch2004} and a spin-only moment of 2.21(2)$\mu_{\rm B}$ determined from high energy photon diffraction.\cite{Strempfer2004} We find the largest perturbation of the Co spin-only moment due to the presence of the \emph{diamagnetic} muon to be about $\pm0.5\%$ and therefore negligible. We believe that the perturbation of the total moment will be similar and therefore have a negligible effect on experimental \musr\ spectra.\cite{supplemental}

In a \musr\ experiment, a muon at position ${\bf r}_{\mu}$ couples to the dipolar field\cite{supplemental} ${\bf B}_{\rm dip}({\bf r}_{\mu})$ of the host's magnetic moments. There is negligible contact hyperfine coupling for the diamagnetic \fmuf\ muons, so in CoF$_2$ the \fmuf\ muons only probe ${\bf B}_{\rm dip}({\bf r}_{\mu})$, which we have calculated for both an unperturbed crystal and an aperiodic perturbed crystal (simulating the presence of the muon).\cite{supplemental} The perturbation of the magnetic moments has been neglected. Our calculations predict a reduction of the dipolar field at the muon site by 21.4\% ($2\times2\times2$ supercell), 23.6\% ($3\times2\times3$ supercell), 23.9\% ($4\times2\times2$ supercell). Experimentally the dipolar coupling is measured to be 16\% lower\cite{Renzi1984Location} than expected from a Co moment of 2.64$\mu_{\rm B}$. This is in reasonable agreement with our prediction and demonstrates that relaxed geometries obtained from DFT are suitable for calculating corrections to expected dipolar fields and hence magnetic moments measured by \musr. Note that the perturbation of the n.n.\ cations in CoF$_2$ is fairly moderate in comparison with the other compounds in this series (Fig.~\ref{fig:distortions}). Nonetheless the n.n.\ cation displacements in CoF$_2$ have a significant effect on the calculated dipolar coupling due to the short-ranged nature of the dipolar interaction.\cite{supplemental} This illustrates that in ionic insulators the muon's perturbation cannot be neglected if magnetic moments are to be measured accurately by \musr. In more covalent compounds we expect the $\mu^+$ charge to be more screened and so structural distortions are probably smaller. 

All of the results above are valid for a positive muon, a proton, or a deuteron defect and are `classical' in the sense that they do not take account of zero-point effects due to the small muon mass ($m_\mu \approx m_{\rm p}/9$). We have used density-functional perturbation theory\cite{baroni2001} (DFPT) to calculate the vibrational properties of the \fmuf\ system in the solid and in vacuum (table~\ref{tab:fmuf}). In vacuum the linear (\fmuf)$^-$ anion has four vibrational modes: symmetric stretch, bending (two-fold degenerate), and asymmetric stretch. In the solid the two-fold degeneracy of the bending mode is broken due to the symmetry of the site: in LiF, NaF, CaF$_2$, and BaF$_2$ there are two fundamentally inequivalent directions of bending; one is towards a neighbouring cation and is shifted up in frequency while the other direction is into a `gap' in the crystal structure and is shifted down in frequency. The asymmetric stretch is very similar to its vacuum value except in CaF$_2$, where the small bond length leads to a larger value. In CaF$_2$, BaF$_2$, and CoF$_2$ all \fmuf\ modes are highly-localized and decouple from the lattice modes. The decoupling of the vibrational modes of the linear \fmuf\ system illustrates that it may be viewed as a molecule-in-a crystal defect similar to the V$_{\rm K}$ center found in the alkali halides.\cite{hayes2004} In LiF and NaF the symmetric stretch mode mixes with the lattice modes so this analogy is slightly less apposite. From the frequencies of the decoupled vibrational modes we have estimated the zero-point energy (ZPE) of the system (in the harmonic approximation). The muon-fluoride (or hydrogen-fluoride) bond is the strongest known hydrogen bond in nature.\cite{hbonding} Combined with the small mass of the muon this leads to the exceptionally large ZPE of the \fmuf\ center of 0.80~eV in vacuum, which is larger than the ZPE of any natural triatomic molecule (the ZPEs of H$_2$O and H$_3^+$ are 0.56 and 0.54 eV respectively\cite{carney1976}). This demonstrates the importance of quantum effects in muon localization.  %(note that for all of the compounds considered here the structural relaxation yields only a single candidate site for dia- and paramagnetic muons). 

We now discuss the properties of the neutral muonium (Mu) state in this series. The calculated equilibrium geometries are shown in Fig.~\ref{fig:structures}. Since the muon charge is screened by an electron, the Mu site is fundamentally different to the diamagnetic site. In LiF and NaF, Mu occupies the octahedral interstitial site. An interstitial site was previously suggested based on the observed hyperfine coupling.\cite{Baumeler1986} In CaF$_2$ and BaF$_2$, Mu occupies the octahedral cation-cation centered site, the fluoride-fluoride centered octahedral site is unstable. In CoF$_2$ we find a single Mu site in a nearly octahedral position at approximately (0.56, 0.84, 0.50), distorted by the neighbouring fluoride. Fig.~\ref{fig:distortions} shows the radial displacements of the ions from the Mu defect. Due to the screening effect of the Mu electron, the displacements are generally much smaller than for the diamagnetic $\mu^+$. In LiF, NaF, CaF$_2$, and BaF$_2$ the displacements of the n.n.\ ions are again along the radial direction due to the symmetry of the Mu site and the ions are only displaced away from the Mu. In CoF$_2$ the symmetry of the Mu site is lower, leading to small displacements in the tangential direction, even for the n.n.\ shell, through elastic interactions with the lattice. These are also the likely cause for the effective attraction of some ions despite the neutral charge state of Mu. All of this also applies to neutral interstitial hydrogen H$_{\rm i}^0$. 

\begin{table}[htbp] %add [H] placement to break table across pages
\begin{ruledtabular}
\begin{tabular}{ l l | l l l l l l l} % left left separator left left left ... 
&  & $A$ & $E_{\rm HA}$ & $\langle A\rangle_{\rm HA}$ & $E_{\rm FD}$ & $\langle A\rangle_{\rm FD}$ & $A_{\rm exp}$ \\ \hline
Vac. & Mu & 4711 & -- & -- & -- & -- & 4463 \\
 & H$_{\rm i}^0$ & 1480 & -- & -- & -- & -- & 1420 \\
LiF & Mu & 4368 & 0.50 & 4256 & 0.51 & 4238 & 4584\cite{Baumeler1986}  \\
  & H$_{\rm i}^0$ & 1372 & 0.18 & 1361 & 0.17 & 1360 & 1400\cite{Kamikawa1980} \\
  NaF & Mu & 4389 & 0.38 & 4293 & 0.42 & 4208 & 4642\cite{Baumeler1986} \\
 & H$_{\rm i}^0$ & 1379 & 0.13 & 1371 & 0.14 & 1367 & 1500\cite{Hoentzsch1979} \\
CaF$_2$ & Mu & 4610 &  0.31 & 4564 & 0.33 & 4564 & 4479\cite{Kiefl1984}  \\
 & H$_{\rm i}^0$ & 1448 & 0.10$^{\rm a}$ & 1440 & 0.10 & 1440 & 1464\cite{Hall1962} \\
BaF$_2$ & Mu & 4605 & 0.20 & 4560 & 0.23 & 4565 & -- \\
 & H$_{\rm i}^0$ & 1447 & 0.07 & 1440 & 0.07 & 1440 & 1424\cite{Hodby1969} \\
CoF$_2$ & Mu & 1281 & 0.62 & 1397 & 0.59 & 1535 & --$^{\rm b}$ \\
 & H$_{\rm i}^0$ & 403 & 0.21 & 420 & 0.20 & 441 & -- \\
\end{tabular}
\end{ruledtabular}
\caption{\label{tab:hf} Calculated and experimental contact hyperfine couplings (MHz) and zero-point energies $E$ (eV) of Mu and H defects. $A$: `classical' hyperfine coupling;  $\langle A\rangle$: quantum corrected value using the harmonic approximation (HA), by solving the full Schr\"odinger equation using finite differences (FD). The ratio of $E_{\rm HA}$ for the Mu and H modes is close to $\sqrt{m_{\rm p}/m_\mu}\approx 3$ indicating highly-localized modes. $^{\rm a}$Measured (Ref. \onlinecite{Shamu1968}) to be 0.12 eV at 100 K. $^{\rm b}$While no experimental value $A_{\rm exp}$ has been reported for CoF$_2$, a total coupling of 1280~MHz has been measured (Ref. \onlinecite{Kiefl1987}) in MnF$_2$. The dipolar coupling (without quantum correction) in CoF$_2$ is $\approx 71$~MHz (0.52~T along $c$) and adds to the contact term.} 
\end{table}

The paramagnetic state is experimentally characterized by the (dipolar and contact) hyperfine coupling between the muon (proton) spin and the surrounding spin density. For all paramagnetic states above, except the one in CoF$_2$, the dipolar coupling cancels by symmetry. Unlike in the diamagnetic case, the n.n.\ Co spin-only moment \emph{is} significantly perturbed ($-25\%$) by the presence of the Mu. The estimate of the dipolar coupling assumes that the relative perturbation of the total moment is also $-25\%$ and takes account of crystallographic distortions.\cite{supplemental} The contact hyperfine coupling $A$ is related to the unpaired spin density $\rho({\bf r}_n)$ at the muon/proton position ${\bf r}_n$ via $A=\frac{2\mu_0}{3}\gamma_e\gamma_n \rho({\bf r}_n)$, where $\gamma_e$ is the electron gyromagnetic ratio and $\gamma_n$ is the muon/proton gyromagnetic ratio. The spin density was obtained using the projector-augmented wave reconstruction method\cite{Blochl1994} and the resulting contact hyperfine couplings are shown in table~\ref{tab:hf}. However, due to the ZPE of the defect, the defect wavefunction has a finite spread leading to a quantum correction to the hyperfine coupling. This has been previously studied in Si, Ge, and diamond.\cite{Porter1999,Luchsinger1997} Although a complete treatment would involve a parameterization of the full three-dimensional contact hyperfine coupling and potential energy surface to calculate the three-dimensional wavefunction,\cite{Porter1999} we obtain an estimate of this correction as follows. The vibrational modes of the defect were calculated using DFPT and the potential energy and hyperfine coupling were calculated along the direction of the eigenmodes. Since the eigenmodes are mutually perpendicular, the motion along the three modes decouples and the wavefunction factorizes. We have then calculated the Mu and H$_{\rm i}^0$ wavefunctions in two ways. The first is an anisotropic harmonic approximation where the wavefunction along each mode is the ground state wavefunction of the harmonic oscillator with the frequency $\omega$ given by the calculated vibrational frequency. Inspection of the potential energy surface has revealed significant anharmonic terms along some directions in some of the compounds.\cite{supplemental} We have therefore also solved the full Schr\"odinger equation along the calculated mode directions using a finite differences method. From the calculated contact hyperfine couplings $A({\bf r})$ and the Mu/H$_{\rm i}^0$ wavefunction $\psi({\bf r})$, we have obtained the estimated quantum correction $\langle A\rangle$ from $\langle A\rangle = \frac{\int r^2 dr |\psi({\bf r})|^2 A({\bf r})}{\int r^2 dr |\psi({\bf r})|^2}$. The calculated couplings $A({\bf r})$ and wavefunction $\psi({\bf r})$ are weighted by $r^2$ to obtain an approximate three-dimensional average. This approximation is accurate so long as $\psi({\bf r})$ and $A({\bf r})$ are approximately spherically symmetric between neighbouring points on the integration grid. Since the average is taken over three mutually perpendicular directions, we expect this to be a reasonable approximation. The quantum corrections are also tabulated in table~\ref{tab:hf}. Note that this correction is smaller for the heavier H$_{\rm i}^0$. Our estimates are within 10\% of the experimental value for LiF and NaF, the same level of accuracy as previous calculations in Si, Ge, and diamond,\cite{Porter1999,Luchsinger1997} and within 2\% of the experimental value for CaF$_2$.

There has been considerable interest recently in identifying muon sites by locating the minima of the electrostatic potential of the unperturbed host.\cite{Luetkens2008,Maeter2009PRB,Bendele2012,deRenzi2012,Prando2013} We have therefore compared the muons sites in this series with the location of the minima of the electrostatic potential of the unperturbed solid, and have found that these do not generally coincide.\cite{supplemental,Bernardini_note} In the diamagnetic case this is primarily due to the formation of the molecular \fmuf\ state. While interstitial Mu interacts more weakly with the host due to the screening by the Mu electron, this screening also makes Mu less sensitive to the host's electrostatic potential and the Mu site is mainly determined by the space required to accommodate the Mu electron. All of the compounds studied here are very ionic in character and the $\mu^+$-lattice interaction is therefore expected to be stronger than in more covalent insulators or metals (where the $\mu^+$ charge would at least be partially screened). Nonetheless we expect the combination of this screening (where operative), the muon-lattice interaction, and the muon's exceptionally large zero-point energy to frequently lead to muon localization away from the minima of the electrostatic potential of the unperturbed host. We therefore believe that muon sites cannot be determined reliably on the basis of the electrostatic potential alone. 

In conclusion we have demonstrated systematically how DFT can be used to comprehensively address the two most fundamental limitations of the \musr\ technique: the problem of the unknown muon site and the perturbation exerted by the muon on its host. We note that the detailed understanding of the nature of the muon's state in solids is relevant beyond the field of \musr\ since the muon acts as a light analogue of hydrogen, which is a ubiquitous impurity in all technologically important semiconductors, where it strongly affects the electronic and structural properties of the material.\cite{VdW2003Nature}

%\begin{acknowledgments}
We thank the following people for useful discussions and technical help: Pietro Bonf\`{a}, Roberto De Renzi, Fabio Bernardini, Nikitas Gidopoulos, Fan Xiao, Jack Wright, Andrew Steele, Andrea Dal Corso, Emine K\"u\c{c}\"ukbenli, Bill Hayes, and Steve Cox. Calculations were performed on computers of the E-Infrastructure South Initiative (UK), the Tr\`{e}s Grand Centre de calcul (France), CINECA (Italy), and EPFL (Switzerland). The muon experiment on CoF$_2$ was performed on the GPS instrument at the Paul-Scherrer Institut, Villigen, Switzerland. This work is supported by EPSRC (UK). 
%\end{acknowledgments}


%%%%%%%%%%%%%%%%% SUPPLEMENT %%%%%%%%%%%%%%%%%

\renewcommand*{\citenumfont}[1]{S#1}
\renewcommand*{\bibnumfmt}[1]{[S#1]}

\renewcommand{\thefigure}{S\arabic{figure}}
\renewcommand{\theequation}{S\arabic{equation}}

\section*{Supplemental Material}

\subsection{Computational details}
The {\it ab initio} calculations were performed with the {\sc Quantum ESPRESSO} package\cite{suppl_qespresso} within the generalized-gradient approximation\cite{suppl_pbe} (GGA) using norm-conserving and ultra-soft\cite{suppl_uspp} pseudopotentials. The functionals did not include the spin-orbit interaction. The muon was modelled by a norm-conserving hydrogen pseudopotential. Total energies were converged to at least $2\times10^{-12}$~Ry/atom (Ry is the Rydberg constant). Structural relaxations, total energies, band structures, and vibrational modes were calculated with wavefunction and charge density cutoffs of 80 and 320 Ry, respectively. Contact hyperfine couplings were calculated using the projector-augmented wave (PAW) method\cite{suppl_Blochl1994} as implemented in the GIPAW package.\cite{suppl_qespresso} The PAW calculation required norm-conserving datasets with wavefunction and charge density cutoffs of 120~Ry and 480~Ry, respectively. The PAW datasets and cutoffs were also used for the calculation of the L\"owdin charges used to estimate magnetic moments. We used a Gaussian broadening of the occupations of 0.01~Ry. Vacuum vibrational modes and hyperfine couplings were calculated in a $20 \times 20 \times 20$~\AA$^3$ cell (at $\Gamma$). The convergence thresholds for the structural relaxation were: $10^{-3}$ Ry/$a_0$ for the forces ($a_0$ is the Bohr radius) and $10^{-4}$ Ry for the energy change between subsequent steps. A uniform negative background charge was used to neutralize the muon's charge for the diamagnetic sites since charged supercells cannot be treated with periodic boundary conditions. 

The calculated lattice parameters for the bulk compounds are within 2\% of the experimental values. The experimental values were used for subsequent calculations. The unit cell was fixed during the structural relaxations which included the muon. By symmetry the ions cannot relax from their experimental positions in the bulk for LiF, NaF, CaF$_2$ and BaF$_2$. In CoF$_2$ we found a small relaxation of the fluorides by 0.037~\AA\ with respect to their experimental positions [DFT: F site $(0.3024,0.3024,0)$, experimental F site: $(0.308,0.308,0)$]. Figs.~1 and 2 in the main text use the calculated fluoride positions for the bulk as reference. 

The considerable reduction of the spin-only moment in CoF$_2$ from the ideal $S=3/2$ value of 3$\mu_{\rm B}$ has been related\cite{suppl_Jauch2004} to the relative size of the $E$ and $D$ anisotropy constants, where the anisotropy term is of the form $DS_z^2+E(S_x^2-S_y^2)$. Crystalline anisotropies are not reproduced accurately without explicit inclusion of spin-orbit coupling in the GGA functional. This is the likely cause for the overestimate of the spin-only moment in our calculations [calculated: $2.68\mu_{\rm B}$ per Co ion, high-energy photon diffraction:\cite{suppl_Strempfer2004} 2.21(2)$\mu_{\rm B}$]. A previous DFT study\cite{suppl_Dufek1994} using the GGA has found a similar spin-only moment of 2.62$\mu_{\rm B}$. Our calculations also predict a spin-only moment of 0.026$\mu_{\rm B}$ on each fluoride ion that experiences a perturbation of up to $\pm0.017\mu_{\rm B}$ due to the muon. Since the fluorine moments are very small, they have been neglected in the calculation of the dipolar coupling. 

The dipolar interaction of the muon at ${\bf r}_{\mu}$ with the magnetic moment ${\bf m}_i$ (assumed to be completely localized) of ion $i$ at position ${\bf r}_i$ is given by  $B_{\rm dip}^{\alpha}({\bf r}_{\mu})=\sum_i D^{\alpha \beta}_i ({\bf r}_{\mu}) m_i^\beta$, where $D_i^{\alpha \beta}=\frac{\mu_0}{4\pi R_i^3}(\frac{3R_i^\alpha R_i^\beta}{R_i^2}-\delta^{\alpha \beta})$ is the dipolar tensor with ${\bf R}_i=(R_i^x,R_i^y,R_i^z)={\bf r}_\mu-{\bf r}_i$. In the calculation of the dipolar coupling in the aperiodic perturbed crystal, both for the diamagnetic and the paramagnetic state, the perturbation of the crystal structure due to the muon was included within one supercell around the muon (with the muon at the center), surrounded by unperturbed ions within a radius of 23~$a$ lattice parameters. 

In general, the magnetic coupling of the muon is a sum of dipolar coupling, contact hyperfine interaction, demagnetization and Lorentz fields. The calculated contact hyperfine coupling for the diamagnetic \fmuf\ muon in CoF$_2$ is $< 1$~MHz and is therefore negligible in comparison with the dipolar coupling. In an antiferromagnet the demagnetizing and Lorentz fields are zero. Hence for the diamagnetic \fmuf\ muon only the dipolar coupling needs to be considered. The short-ranged nature of the dipolar coupling can be illustrated by separately considering the dipolar coupling of the \fmuf\ muon with the nearest neighbour (n.n.) and the remaining cations in the crystal: assuming a Co moment of 2.64$\mu_{\rm B}$, the crystallographic distortion reduced the dipolar field from the two n.n.\ Co ions from $-0.378$ to $-0.309$~T, while the dipolar field from all other ions in the solid merely changes from 0.113 to 0.101~T (all along the c-axis and for the $2\times2\times2$ supercell). If the cations in the non-magnetic compounds LiF, NaF, CaF$_2$, and BaF$_2$ were magnetic and aligned antiferromagnetically (ferromagnetically) along the c-axis, the dipolar corrections for the \fmuf\ states shown in Fig.~1 in the main text would be: LiF -39\%(-70\%), NaF -34\%(-67\%), CaF$_2$ 0\%(-46\%), and BaF$_2$ 0\%(-47\%). In the antiferromagnetic case, the muon would be located in a site of cancellation of the dipolar field in CaF$_2$ and BaF$_2$. 

\begin{figure*}[htbp]
\includegraphics[width=0.95\textwidth]{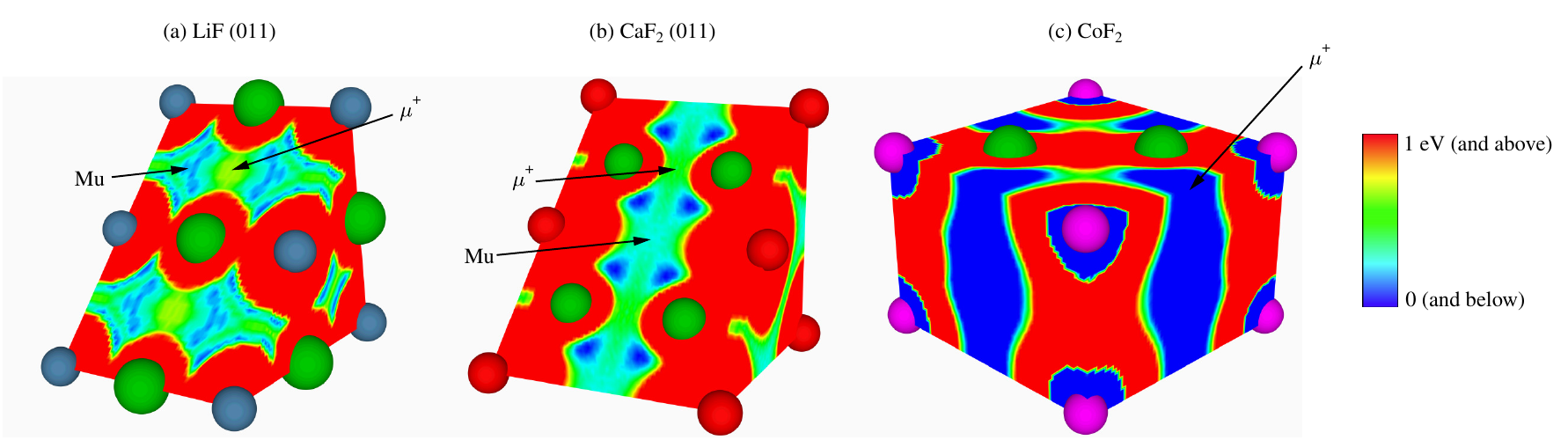}
\caption{\label{fig:potential}(color online). Calculated electrostatic potential for the unperturbed solid. Blue coloring indicates regions that are attractive to a positive charge, red regions repel a positive charge. Below and above the end of the scale the color coding is blue and red, respectively, with no further gradient. The scale is relative and cannot be compared between different compounds. Ions are drawn at their ionic radii. Li (blue), F (green), Ca (red), Co (magenta). The c axis is vertical. Arrows indicate the dia- and paramagnetic muon sites obtained through a full relaxation, which agree with the experimentally determined muon sites. In CoF$_2$ the muonium site is close to the octahedral site that also hosts the diamagnetic muon. Note also that the muon zero point energy, characterizing the extent of its delocalization in the absence of bonding, is about 0.8~eV in the \fmuf\ state and about $0.2-0.6$~eV as muonium, see tables I and II in the main text. The data were visualized with {\it VESTA}.~\cite{suppl_Momma2011}}
\end{figure*}

For Mu in CoF$_2$ the dipolar coupling was estimated from the dipolar coupling to the Co moments only. The Mu electron spin density is approximately spherically symmetric and therefore only yields a small contribution to the dipolar coupling, which has been neglected. The Co moment was assumed to be 2.64$\mu_{\rm B}$ and the perturbation of the n.n.\ moment was taken to be $-25\%$ (the calculated reduction of the spin-only moment). The perturbation of the spin-only moments of the other Co ions in the supercell was negligible. At different levels of approximation the dipolar coupling is 0.49~T (unperturbed crystal), 0.72~T (crystallographic distortions only) and 0.52~T (crystallographic distortions and perturbation of the n.n.\ Co moment), all are along $c$ and have the same sign as the contact coupling. 

\subsection{Comparison with electrostatic potential}
There has been considerable interest recently in identifying muon sites by locating the minima of the electrostatic potential of the unperturbed host (calculated at varying levels of complexity).\cite{suppl_Luetkens2008,suppl_Maeter2009PRB,suppl_Bendele2012,suppl_deRenzi2012,suppl_Prando2013} In this section we compare the sites of the dia- and paramagnetic muons obtained through a full ionic relaxation (which, as demonstrated, are in excellent agreement with the experimental sites) with the location of the minima of the electrostatic potential of the unperturbed solid. We define \emph{electrostatic potential} to mean the inverted sum of the conventional Hartree and ionic potentials (conventionally defined to be positive in regions that repel electronic charge density). The calculated electrostatic potentials are shown in Fig.~\ref{fig:potential} for three of the compounds of this series. It is evident that the minima of the electrostatic potential do not coincide in general with the correct dia- or paramagnetic muon sites.\cite{suppl_Bernardini_note} In the diamagnetic case this is due to the interaction of the muon with its host and in particular the formation of the molecular \fmuf\ state which, having the strongest known hydrogen bond,\cite{suppl_hbonding} releases a substantial amount of energy upon formation. While interstitial muonium generally interacts more weakly with the host due to the screening by the Mu electron, this screening also makes muonium less sensitive to the host's electrostatic potential and the site of muonium localization is mainly determined by the space required to accommodate the Mu electron. Muonium could also be located in a bond-centered rather than an interstitial location, in which case there usually is a significant interaction with the lattice.\cite{suppl_Porter1999,suppl_Luchsinger1997} As argued in the main text, the muon can have an exceptionally large zero-point energy which needs to be taken into account if different candidate sites are investigated (in this series the ionic relaxation only yielded a single dia- and paramagnetic site). It is therefore clear that muon sites should not be assigned to the minima of the electrostatic potential of the unperturbed solid without detailed analysis. 

\subsection{Experimental search for \fmuf\ state in cobalt(II) fluoride}
\begin{figure}[htbp]
\includegraphics[width=0.33\textwidth]{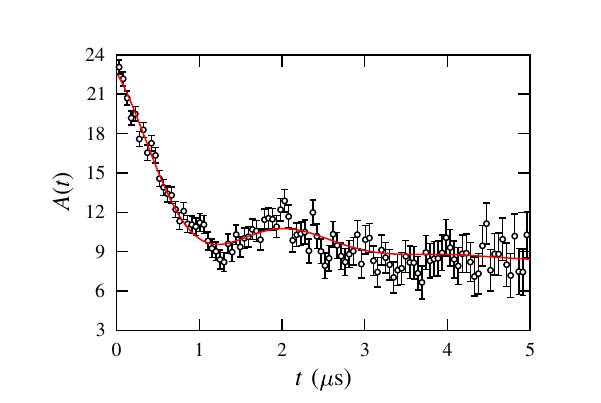}
\caption{\label{fig:cof2}(color online). Experimental muon decay asymmetry $A(t)$ at 70~K in CoF$_2$. The red line is the fit to Eq.~\ref{eqn:fuf}.}
\end{figure}
A powder sample of CoF$_2$ (Sigma Aldrich 236128) was wrapped in 25~$\mu$m silver foil and mounted in a $^4$He cryostat on the GPS instrument at the Paul Scherrer Institut in Switzerland. Above the critical temperature of 37.85~K, we observed oscillations in the muon decay asymmetry $A(t)$ characteristic of an \fmuf\ state, see Fig.~\ref{fig:cof2}. The data were fitted to 
\begin{equation}
A(t)=A_1 \exp(-\lambda_1 t) D_{z}(t)+A_{\rm bg}\exp(-\lambda_{\rm bg} t),
\label{eqn:fuf}
\end{equation} 
where $D_{z}(t)$ describes the time evolution of the muon spin in an \fmuf\ state\cite{suppl_Lancaster2007PRL} and the second term accounts for a slow-relaxing background due to muons stopping in the cryostat tail or sample holder. The term $\exp(-\lambda_1 t)$ phenomenologically takes account of residual magnetic dynamics in the sample with a fitted value of $\lambda_1=0.73(8)$~MHz at 70~K. The best fit is for a symmetric, linear \fmuf\ state with a fluoride-fluoride separation of 2.43(2)~\AA\ in good agreement with our calculated value of 2.36~\AA. The extracted \fmuf\ geometry is not very sensitive to the precise choice of parameterization in Eq.~\ref{eqn:fuf} as it is dominated by the nature of the function $D_z(t)$ which encodes this geometry. 

\subsection{Quantum correction of contact hyperfine coupling}
\begin{figure}[htbp]
\includegraphics[width=0.49\textwidth]{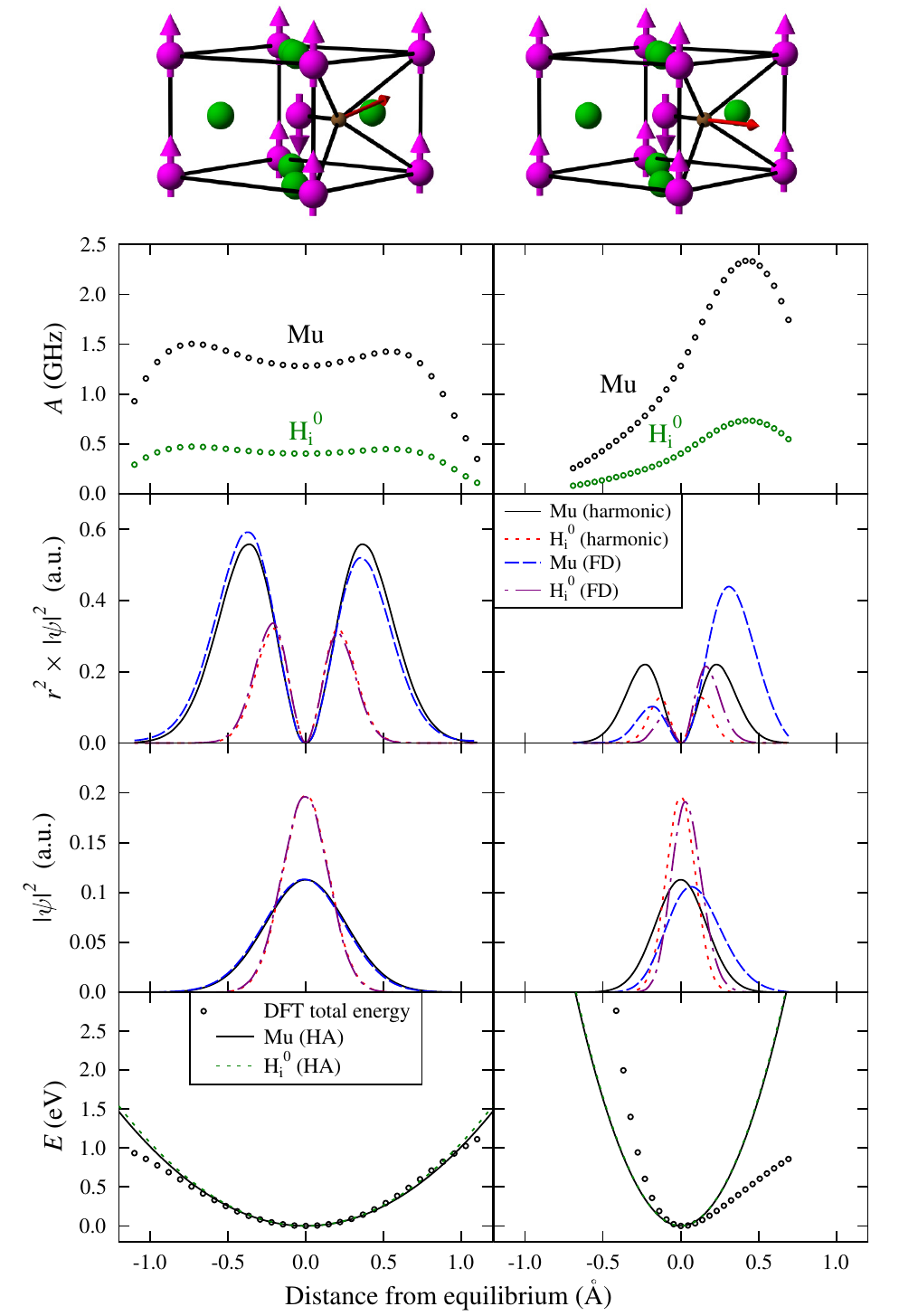}
\caption{\label{fig:qav}(color online). Quantum correction to contact hyperfine coupling using both the harmonic approximation (HA) and by solving the full Schr\"odinger equation using finite differences (FD) along two eigenmodes in CoF$_2$. From top to bottom: schematic of the eigenmode, the red arrow indicates displacement in positive direction; the contact hyperfine coupling for Mu (black) and H$_{\rm i}^0$ (green); the wavefunction weighted by $r^2$; the normalized unweighted wavefunction for Mu (HA black; FD blue) and H$_{\rm i}^0$ (HA red; FD purple); the potential energy above equilibrium calculated from the DFT total energy and the harmonic approximation to it for Mu and H$_{\rm i}^0$. While for the mode on the left the HA is quite reasonable, there are large deviations for the second mode.}
\end{figure}

In the anisotropic harmonic approximation, the ground state Mu (H$_{\rm i}^0$) wavefunction is given by
\begin{equation}
\psi_{\rm HA}({\bf r})  = \ \left ( \frac{1}{R_1^2 R_2^2 R_3^2 \pi^3}\right )^{1/4}  e^{-(r_1'^2+r_2'^2+r_3'^2)/2},
\label{eqn:harmonic}
\end{equation}
where $r_i'=r_i/R_i$ is the distance from equilibrium along eigenmode $i$ of angular frequency $\omega_i$, normalized by the `range' of the wavefunction $R_i =\sqrt{\frac{\hbar}{m \omega_i}}$ and $m$ is the mass of the particle. Along each mode we have calculated the contact hyperfine coupling $A$ and the total energy up to 3~$R_i$ from equilibrium in positive and negative direction at equal spacings. For the finite differences method, we have used the total energies from the supercell calculation as the potential for which the Schr\"odinger equation was solved. The quantum average described in the main text is then found from 
\begin{equation}
\langle A\rangle = \frac{\displaystyle \sum_{i,j} (r_i^j)^2 |\psi(r_i^j)|^2 A(r_i^j)}{\displaystyle \sum_{i,j} (r_i^j)^2 |\psi(r_i^j)|^2},
\label{eqn:qavimpl}
\end{equation}
where $i$ sums over modes and $j$ is the step along each eigenmode. An example calculation is shown in Fig.~\ref{fig:qav}.

\end{document}